\documentclass[letterpaper,12pt]{article}
\usepackage{amsmath}
\usepackage{graphicx,psfrag,epsf}
\usepackage{enumerate}
\usepackage{natbib}
\usepackage{textcomp}
\usepackage[hyphens]{url} % not crucial - just used below for the URL
\usepackage{hyperref}
\usepackage{lmodern}
\usepackage[mathscr]{euscript}
\usepackage{color}

\usepackage{amssymb}
\usepackage{comment}

\usepackage{mathptmx}

\usepackage{graphics}
\usepackage{amsthm}
\usepackage{amssymb}
\usepackage{bm}

\begin{document}

\def\spacingset#1{\renewcommand{\baselinestretch}%
{#1}\small\normalsize} \spacingset{1}

\title{\bf Going Deep:  Models for Continuous-Time Within-Play Valuation of Game Outcomes in American Football with Tracking Data}

  \author{
    Ronald Yurko, Francesca Matano, Lee F. Richardson, \\
    Nicholas Granered, Taylor Pospisil, Konstantinos Pelechrinis, \\
    Samuel L. Ventura}

\maketitle

%% Do NOT include any frontmatter information; including the title, author names,
%% institutes, acknowledgments and title footnotes (author information, funding
%% sources, etc.). Start the document with the first section or paragraph of
%% the article.
\bigskip

\begin{abstract}
Continuous-time assessments of game outcomes in sports have become increasingly common in the last decade. In American football, only discrete-time estimates of play value were possible, since the most advanced public football datasets were recorded at the play-by-play level. While measures such as expected points and win probability are useful for evaluating football plays and game situations, there has been no research into how these values change throughout the course of a play.  In this work, we make two main contributions:  First, we introduce a general framework for continuous-time within-play valuation in the National Football League using player-tracking data. Our modular framework incorporates several modular sub-models, to easily incorporate recent work involving player tracking data in football.  Second, we use a long short-term memory recurrent neural network to construct a ball-carrier model to estimate how many yards the ball-carrier is expected to gain from their current position, conditional on the locations and trajectories of the ball-carrier, their teammates and opponents. Additionally, we demonstrate an extension with conditional density estimation so that the expectation of any measure of play value can be calculated in continuous-time, which was never before possible at such a granular level.
\end{abstract}

\noindent%

\section{Introduction}
\label{sec:intro}
Quantitative analyses of sports have become increasingly complex in the last decade, mostly due to the advent of player and object tracking data across most major sports. Tracking data captures the position and trajectory of the athletes and objects of interest (e.g. balls, pucks, etc) on the playing surface for a given sport. Statistical analysis of tracking data in sports has been an increasingly popular area of research in recent years; we encourage interested readers to read the review paper on this topic from \cite{GudmundssonH16} for a detailed summary of the work in this area.

\begin{figure}[!ht]
    \centering
    \includegraphics[width= \textwidth, height=10cm]{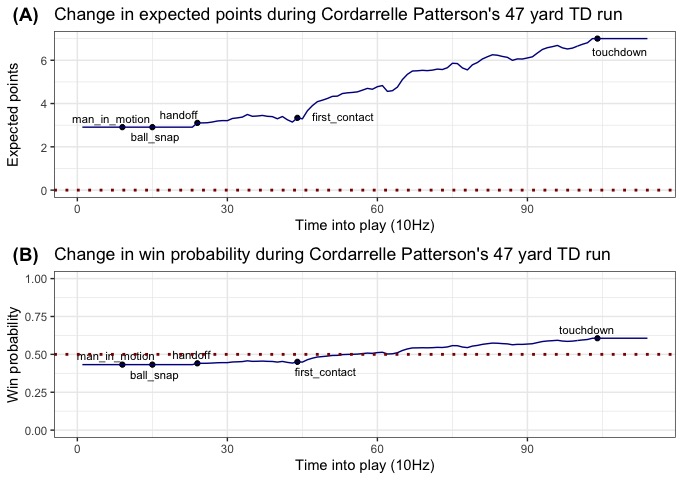}
    \caption{The change in (A) expected points and (B) win probability during Cordarrelle Patterson's 47-yard touchdown run based on random forests for conditional density estimation of the end-of-play yard line.}
    \label{fig:ep-wp-example}
\end{figure}

In this work, we focus on a particular but important area of player tracking data analysis: continuous-time valuation of game outcomes -- in our case, for American football. Figure \ref{fig:ep-wp-example} provides a visual representation of this idea, showing how the expected points (A) and win probability (B) change continuously in reaction to on-field events throughout the course of a fourty-seven yard touchdown run by Cordarrelle Patterson.

Below, we provide a brief overview of discrete-time valuation of game outcomes in football, continuous-time valuation of game outcomes in all sports, and continuous-time valuation of game outcomes in football specifically.

\subsection{Previous Work:  Discrete-Time (Play-by-Play) Evaluation of Football Game Outcomes}
\label{sec:prev-play}

Commonly, there are two classes of models for discrete-time evaluation of game outcomes in football:  {\it expected points} (EP) and {\it win probability} (WP).  Models for EP seek to answer the question:  How many points is the current game situation worth, in expectation, conditional on the features of that game situation (e.g. down, distance, yard line, score differential, time remaining, etc)?  Models for WP ask a fundamentally different question:  How likely is it that the possession team will win the game, conditional on the features of that game situation (e.g. down, distance, yard line, score differential, time remaining, etc)?  \cite{Yurko19} provide an overview of these play-valuation frameworks, including a review of prior approaches for building these models, new approaches for building these models that are publicly available via the \texttt{nflscrapR R} package \citep{Horowitz17, R17}, and examples of how these models and their derived metrics can be used to evaluate individual players and teams. These models are typically estimated at the play-by-play level (\emph{between plays}), since this is the finest level of granularity at which datasets are available. However, there has been no work to-date studying how valuation of football game outcomes evolves \emph{within plays}.

\subsection{Previous Work:  Continuous-Time Models for Game Outcomes in Sports}
\label{sec:prev-sports}

Although tracking data is not \emph{technically} collected in continuous-time -- most systems track the locations and trajectories of athletes and objects of interest at rates of 10 to 25 Hz -- it is fundamentally different from play-by-play or event-level datasets.  In particular, the unit of interest in play-by-play or event-level data is a single (discrete) play or event, while the units of interest in tracking data are the continuously changing locations and trajectories of players and objects on the playing surface.

Using tracking data, several approaches exist for continuous-time modeling of game outcomes in sports. In basketball, \cite{Cervone14} and \cite{Cervone16} provide models for {\it expected possession value} (EPV), which is a continuous-time estimate of the expected points scored by the team in possession during a single basketball possession, conditional on the locations and trajectories of players (and the ball).  The authors use a two-level Markov chain approach to do this.  First, they model the competing hazards of (discrete) possession-changing events (e.g. passes, shot attempts, turnovers).  Second, they model (continuous) player movement on the court.  These two models, each of which condition on the locations and trajectories of the players and the ball, are combined hierarchically to estimate EPV at each moment.  

In soccer, \cite{Link16} quantifies the performance of attacking teams in terms of their probability of scoring.  The authors provide continuously updating estimates of the probability of a goal being scored at each moment throughout the course of a possession.  \cite{fernandezdecomposing} use deep learning to estimate EPV in soccer.  They take a multi-level approach similar to \cite{Cervone16}, where discrete-time estimates of ``expected goals'' (describing the likelihood of a shot resulting in a goal, if taken), ``passing value'' (describing the value, in terms of expected goals, of a pass), and ``drive value'' (describing the value, in terms of expected goals, of a drive to the net) are combined with continuous-time estimates of action likelihood (shot, pass, or drive) to provide an overall, continuous-time measure of EPV.  Each of the sub-models in this approach conditions on the locations and trajectories of the players and the ball.

Observant readers will note several similarities between our framework and the approaches of \cite{Cervone16} and \cite{fernandezdecomposing}:  combining discrete-time and continuous-time models, continuously estimating the value of game outcomes within plays, and using the resulting metrics to quantify the value added of individual athletes.

\subsection{Previous Work:  Continuous-Time Models for Football}
\label{sec:prev-football}

In December 2018, the National Football League (NFL) temporarily made public a subset of player and ball-tracking data from the first six weeks of the 2017 regular season for its inaugural ``Big Data Bowl'' competition.  Although the data has since been taken down, several authors have contributed interesting work to the literature using this data.

\cite{Burke19} introduced a deep learning approach, called DeepQB, to model outcomes of the passing game.\footnote{DeepQB was not developed using the ``Big Data Bowl'' sample but rather every pass attempt from every regular/post-season game in the 2016 and 2017 seasons.} In different variants of this model, the author uses DeepQB to model each receiver's target probability, the pass outcome probability (complete, incomplete, interception), and the expected yards gained.  Each of these variants of DeepQB can be incorporated into the general framework for within-play valuation of game outcomes that we provide in this paper.

\cite{Deshpande19} provide innovative statistical models for the hypothetical completion probability of a pass.  The authors use counterfactual analysis of within-play features to impute upstream and downstream features like the time at which the ball will arrive to the targeted receiver.  This model can also be incorporated into the general framework for within-play valuation of game outcomes that we provide in this paper.

Several other authors have undertaken interesting research topics using the NFL-provided tracking data.  For example, \cite{Chu19} use mixture modeling to automatically identify, cluster, and characterize route types of receivers.   Similarly, \cite{Sterken19} use a convolutional neural network to classify the route types of receivers. \cite{Dutta19} use clustering models to provide unsupervised, probabilistic annotations for the coverage type of defensive backs.
\cite{VonderHaar19} provides an exploratory analysis of NFL passing plays. These works all involve improving upon the existing league-provided tracking data by providing additional information that can be estimated from the underlying player locations and trajectories.  However, they do not attempt to model game outcomes, so they are of limited relevance to this paper.

\subsection{Our Contributions}

Our paper makes two main contributions.  First, we provide a general framework for continuous-time within-play valuation of game outcomes in the NFL, using the league-provided tracking data.  Our framework, described in Section \ref{sec:framework}, incorporates several modular sub-models, so that the recent work involving player tracking data in football described above can be easily incorporated into our framework.  

Second, we construct a novel {\it ball-carrier model}, which estimates the expected yards gained from a ball-carrier's current position (and thus, the end-of-play yard line), conditional on the locations and trajectories of the ball-carrier, their teammates, and their opponents. We focus on modeling the continuous-time end-of-play yard line in this manuscript because the between-play estimates, EP and WP, are essentially functions of the end-of-play yard line, which determines the down, yards to go, possession team, and so on. We find that long short-term memory (LSTM) recurrent neural networks outperform alternative approaches for this modeling task. By continuously updating the end-of-play yard line predictions from the LSTM at each frame of the tracking data, we can evaluate ball-carrier performance within plays (examples provided in Sections \ref{sec:ep-play-ex} and \ref{sec:ep-players}). Finally, we demonstrate an extension to our ball-carrier model using conditional density estimation in Section \ref{sec:discussion}, from which we can compute the continuous-time within-play EP and WP estimates in Figure \ref{fig:ep-wp-example}.

Our research has several key benefits:  First, the framework is adaptable, so that measure of play value (or any model for EP or WP) can be used.  Second, the framework is modular, so that (for example) any model for pass attempt outcomes or quarterback decision-making can be substituted into this framework in place of the approach we use here.  For example, one could use the models from \cite{Burke19} or \cite{Deshpande19} in the appropriate places of the framework described in Section \ref{sec:framework}.  Finally, although our focus on player evaluation is limited in this paper, the fully-implemented framework will allow for continuous-time assessment of off-ball player movement, quarterback decision-making, ball-carrier value added, receiver value added, blocking value added, defensive player value added, and many other evaluative tools that were never before possible at such a granular level.

\section{Player and Ball Tracking Data}
\label{sec:data}

In December 2018, the NFL became the first North American professional sports league to release a portion of their tracking data to the public when temporarily made available a subset of this data from the first six weeks of the 2017 season for the inaugural ``Big Data Bowl'' competition.\footnote{The NFL ran a separate competition involving analyzing tracking data for punts, but since it only covered punt plays, it is not relevant for this paper.}  

The NFL's tracking data collected as follows:  Two radio frequency identification (RFID) chips are placed in each player's shoulder pads (and in the ball).  The RFID chips emit a signal to sensors in each stadium, which triangulate the location of the chip on the field.  The data is collected at a rate of 10 Hz, so that the on-field location, speed, and angle of each player (and the ball) is recorded 10 times per second.  Event annotations (e.g. ball snapped, first contact, pass thrown, etc) are recorded by the NFL for each play.  In total, the dataset contains 1,075,720 unique frames across 14,167 plays, each of which records the locations and trajectories (speed, angle) of all twenty-two players (and the ball) on the field.  

\begin{figure}[!ht]
    \centering
    \includegraphics[width= \textwidth]{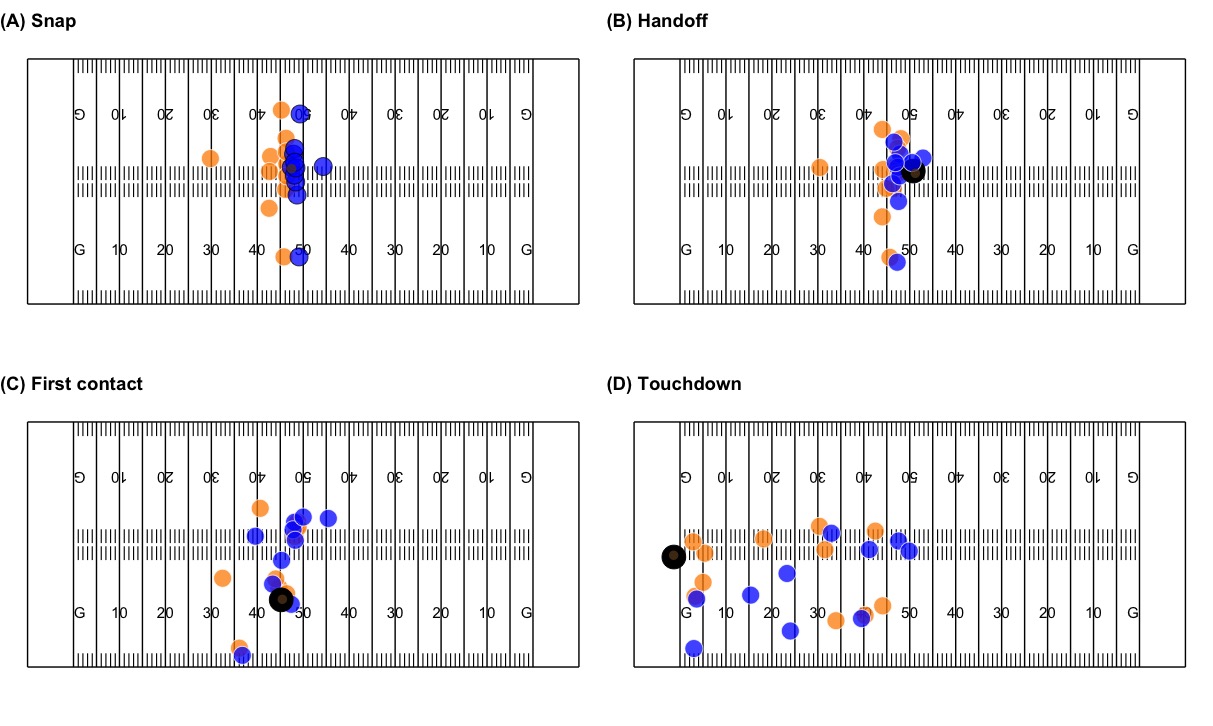}
    \caption{A display of the tracking data for Cordarrelle Patterson's 47-yard TD run with the offense (blue), defense (orange), and ball-carrier (black) at (A) snap, (B) handoff, (C) first contact, and (D) crossing the endzone.}
    \label{fig:explay}
\end{figure}

Table \ref{tab:exdata} shows an example of this data for a 47-yard touchdown run by WR Cordarrelle Patterson, which occurred in a Week 6 game between the Los Angeles Chargers and Oakland Raiders in the 2017 season. Four frames from this play are displayed in \autoref{fig:explay} displaying the coordinates of the offense (blue), defense (orange), and the ball-carrier (black) at particular events in the play. We will visualize the player tracking data in this manner for the remainder of the manuscript.

\begin{table}[!ht]
\centering
\caption{Example of tracking data for Cordarrelle Patterson's 47-yard TD run.}
\label{tab:exdata}
\centering
    \begin{tabular}{|c|c|c|c|c|c|c|}
        \hline
        frame.id & x & y & s & dir & event & displayName   \\
        \hline
        24 & 60.64 & 29.70 & 7.55 & 175.34 & handoff & Cordarrelle Patterson	  \\
        \hline
        25 & 60.77 & 28.94 & 7.61 & 177.10 & NA & Cordarrelle Patterson   \\
        \hline
        \vdots & \vdots & \vdots & \vdots & \vdots & \vdots & \vdots   \\
        \hline
        44 & 55.20 & 14.62 & 8.92 & 226.45 & first\_contact & Cordarrelle Patterson  \\
        \hline
                \vdots & \vdots & \vdots & \vdots & \vdots & \vdots & \vdots   \\
    \end{tabular}
\end{table}

\begin{figure}[!ht]
    \centering
    \includegraphics[width= \textwidth, height=10cm]{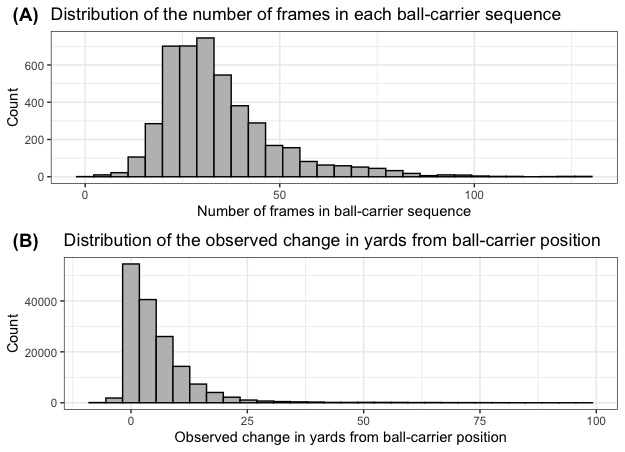}
    \caption{Distributions of the (A) length of the ball-carrier sequences in the modeling dataset, and (B) the observed change in yards from the ball-carrier's location at the current frame with respect to the target endzone.}
    \label{fig:bc-hist}
\end{figure}

\noindent This data can easily be joined to existing play-by-play data from the NFL's API (e.g. via the {\tt nflscrapR} package), which contains additional information about each play \citep{Horowitz17}.  
For the models in Section \ref{sec:ball-carrier}, we identified all ball-carrier sequences for running plays, which includes designed runs and QB scrambles. While the tracking data records the location of the ball in addition to the players, it does not identify who is the ball-carrier for a particular frame. We first identified the ball-carriers for every type of play (pass attempts, runs, returns, etc.) based on the information available from the NFL's API via \texttt{nflscrapR}, which denotes who was directly involved in each play. Given the roles a player can have (passer, runner, receiver, interceptor, or returner), we used the provided event annotations to determine when a player became the ball-carrier. Since, for simplicity, we focus our attention on running plays in this manuscript, we identify the beginning of the ball-carrier sequence when the runner received the ball by either a handoff, lateral, or direct snap. The end of the ball-carrier sequence was marked when either the player was tackled, ran out of bounds, fumbled, or scored a touchdown. We excluded all plays missing the necessary information from the NFL API, as well as plays where the snap of the play was missing in the tracking data, and any ball-carrier sequences where either the starting or ending events were missing. After further pre-processing for the covariates described in Section \ref{sec:ball-carrier}, our final modeling dataset consisted of 154,908 frames from 4,502 unique ball-carrier sequences on running plays. \autoref{fig:bc-hist}(A) displays the distribution of the length of these ball-carrier sequences, revealing that majority of ball-carrier sequences are between two to five seconds in length, while \autoref{fig:bc-hist}(B) displays the observed change in field position from the ball-carrier's current location that will be modeled, as discussed in \ref{sec:ball-carrier-intro}.

\section{A Framework for Continuous-Time Play Value in Football}
\label{sec:framework}

Our approach for providing continuous-time within-play valuations involves several key pieces, with the main framework presented in  Section \ref{sec:framework-detail}. We first describe 
several sub-models for computing various within-play quantities that comprise our within-play valuation framework:  A ball-carrier model (Section \ref{sec:ball-carrier-intro}), a quarterback decision model (Section \ref{sec:qb-decision}), a target probability model (Section \ref{sec:target-prob}), a global catch probability model (Section \ref{sec:incompletion-model}), and an individual catch probability model (Section \ref{sec:catch-model}). Since the primary determinant of the inputs to models for football game outcomes is the end-of-play yard line, we describe our framework in terms of predicting the expected end-of-play yard line. We highlight the modularity of our framework in Section \ref{sec:discussion} with a replacement for the ball-carrier model that, when combined with models for evaluating game situations at a discrete level between each play, yields continuous-time valuation of game outcomes for American football, as demonstrated in Figure \ref{fig:ep-wp-example}.

\subsection{Notation}
\label{sec:notation}

Here, we summarize the notation used in the rest of this section, for easy reference:

\begin{itemize}
    \item $t>0$ is some time between the start (i.e. the snap) and end of a play,
    \item $Y$ is a random variable representing the yards gained from the ball-carrier's current position on the field, and $Y^*$ is the corresponding end-of-play yard line,
    \item $X_t$ is a data structure representing the locations and trajectories of all players and the ball from the start of the play until time $t$,
    \item $\mathscr{F}(X_{t,i})$ is some filtration of the locations and trajectories of all players and the ball from the start of the play until time $t$ on play $i$, borrowing notation from \cite{Cervone16},
    \item $\mathbb{E}[Y_{t,i}|\mathscr{F}(X_{t,i})]$ is the expected yards gained from the ball-carrier's current position, and $\mathbb{E}[Y_{t,i}^*|\mathscr{F}(X_{t,i})]$ is the corresponding expected end-of-play yard line,
    \item $T_{j,i}$ is a binary random variable describing whether receiver $j$ was targeted or not on play $i$ ,
    \item $C_{i}$ is a binary random variable describing whether a pass is caught or not by an offensive or defensive player on play $i$,
    \item $C_{k,i}$ is a binary random variable describing whether player $k$ caught the ball or not on play $i$, where $k$ represents one of the 16 players who can catch a pass (five eligible offensive receivers and 11 defenders),
    \item $P(D_i = d_k|\mathscr{F}(X_{t,i}))$ is a probability mass function over the set of decisions a QB can make:  $\{d_{1} = \mbox{throw away}, d_{2} = \mbox{run/sack}, d_{3} = \mbox{pass}\}$,
    \item $P(T_{j,i}|\mathscr{F}(X_{t,i}), D_i = \mbox{Pass})$ is a probability mass function describing the likelihood that a receiver is targeted on play $i$,
    \item $P(C_i | \mathscr{F}(X_{t,i}), D_i = \mbox{Pass}, T_j = 1)$ is a probability mass function describing the outcome (catch or no catch) of a pass on play $i$ targeted to receiver $j$,
    \item $P(C_{k,i} | \mathscr{F}(X_{t,i}), D_i = \mbox{Pass}, T_j = 1, C_i = 1)$ is a probability mass function describing whether player $k$ caught the ball or not.
\end{itemize}

\subsection{Framework for Continuous-Time Modeling in American Football}
\label{sec:framework-detail}

As the first step towards building a continuous-time valuation framework for American football, we model the expected end-of-play yard line. Our framework for providing continuously-updating within-play valuations is organized as follows: 

\vskip 2 mm

\noindent \textbf{Rushing Plays}: Model the expected yards gained from the ball-carrier's current position, $\mathbb{E}[Y_{t,i}|\mathscr{F}(X_{t,i})]$. Then obtain the associated expected end-of-play yard line through linearity of expectations,
\begin{equation}
    \label{eq:exp-yard}
    \mathbb{E}[Y_{t,i}^*|\mathscr{F}(X_{t,i})] = \mathbb{E}[Y_{t,i}|\mathscr{F}(X_{t,i})] + \text{ [player's current yard line]}.
\end{equation}

\vskip 2 mm
    
\noindent \textbf{Dropbacks}:  Model the QB's decision probabilities, $P(D_i|\mathscr{F}(X_{t,i}))$:
    \begin{itemize}
        \item \textbf{$D_i = $ Throw away}: play ends at the play's original yard line,
        \item \textbf{$D_i = $ Scramble or sack}: use ball-carrier model to estimate the expected end-of-play yard line,
        \item \textbf{$D_i = $ Pass}:  Model $P(T_{j,i} = 1|\mathscr{F}(X_{t,i}), D_i = \mbox{Pass})$, each offensive receiver $j$'s target probability on play $i$ (normalize these probabilities at each time $t$)\footnote{We suggest the use of Softmax normalization here, to handle rare cases where the estimated target probabilities are all 0.}
        \begin{itemize}
            \item \textbf{For each offensive receiver $j$}:  Model $P(C_i = 1 | \mathscr{F}(X_{t,i}), D_i = \mbox{Pass}, T_j = 1)$, the \textit{global} catch probability of a pass on play $i$ by any player on either offense or defense:
            \begin{itemize}
                \item \textbf{No catch}:  play ends at the play's original yard line,%; update the covariates for the play value model accordingly (e.g. increment the down, adjust the time remaining, maintain same yard line)
                \item \textbf{Catch (includes interceptions)}:  Model  $P(C_{k,i} = 1 | \mathscr{F}(X_{t,i}), D_i = \mbox{Pass}, T_j = 1, C_i = 1)$, the \textit{individual} catch probability for each potential pass-catcher $k$ (any of the five offensive receivers and eleven defenders; normalize these probabilities to $P(C_i = 1 | \mathscr{F}(X_{t,i}), D_i = \mbox{Pass}, T_j = 1)$):
                \begin{itemize}
                    \item \textbf{For each potential pass-catcher $k$}:  use ball-carrier model to estimate the expected end-of-play yard line.
                \end{itemize}
            \end{itemize}
        \end{itemize}
    \end{itemize}

\begin{figure}[!ht]
    \centering
    \includegraphics[width= \textwidth]{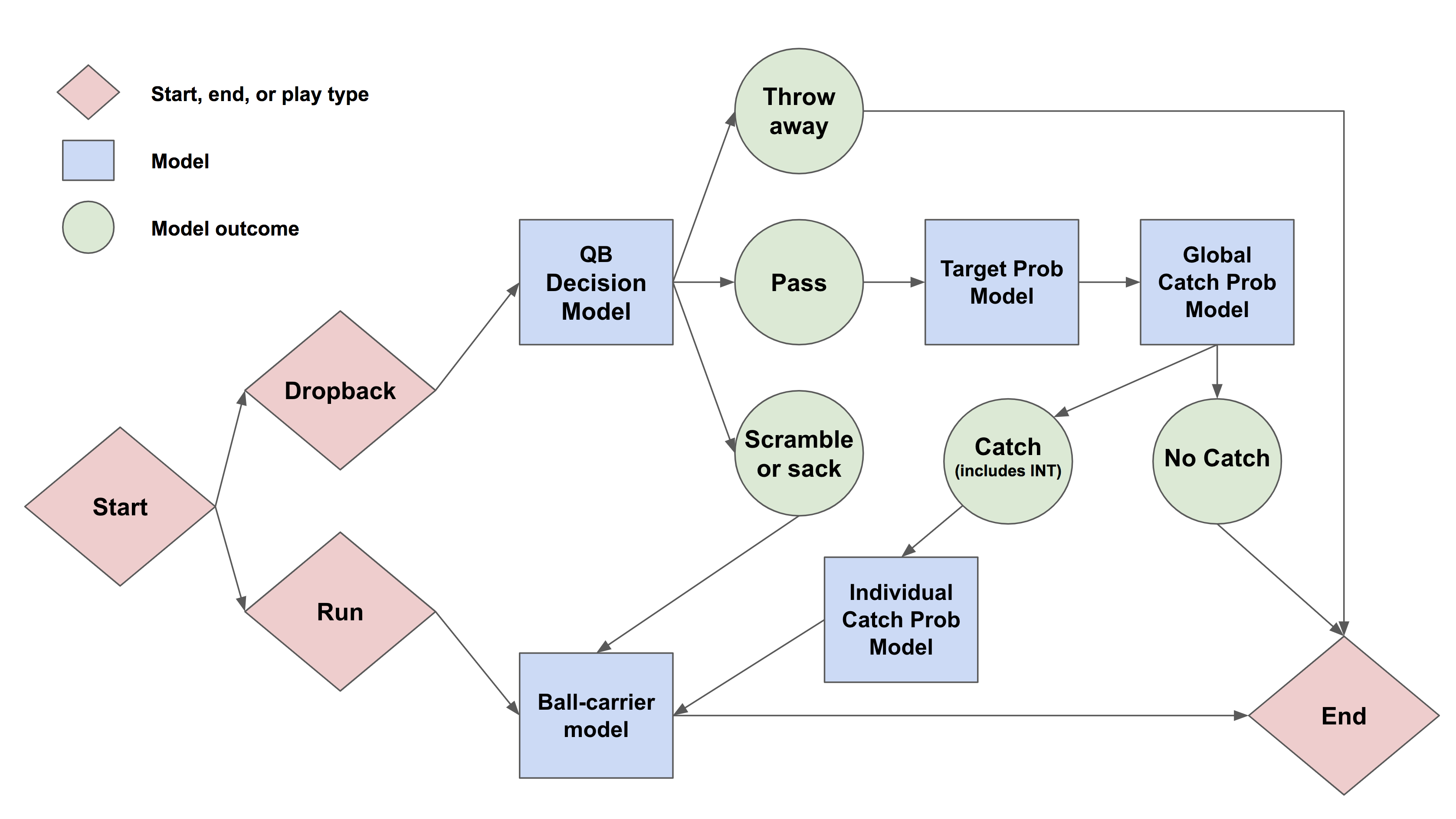}    
    \caption{Continuous-time play value framework. The blue squares represent sub-models, that can be estimated independently. The green circles are discrete outcomes of previous events, and the red diamonds indicate either the start of the play, the end of the play, or whether the play is a run or a pass.}
    \label{fig:framework}
\end{figure}

The above framework is illustrated in Figure \ref{fig:framework}. In the above framework, the predictions from every model are updated at each time $t$ throughout the play, and (given the play type) can be combined to get an overall expected end-of-play yard line.  For rushing plays, the expected end-of-play yard line is directly estimated.  For passing plays, each possible node on the decision tree in the framework above has two pieces of information:

\begin{enumerate}
    \item the node's probability of being achieved, which is computed using the estimated probabilities at each step/split in the tree,
    \item the expected end-of-play yard line, since each node eventually ends with the ball-carrier model's estimate of the expected yards gained (or ends without a ball-carrier, in the case of an incompletion or throw away).
\end{enumerate}

\noindent These two pieces of information are easily combined across all nodes into a single estimate of the expected end-of-play yard line. In this manuscript, we implement the ball-carrier model since it ultimately determines the expected end-of-play yard line (see Section \ref{sec:ball-carrier}). We leave the remaining models of the framework for future work, but describe possible approaches one can take. Additionally, we assume the play type (run or dropback) is known at the start of the play, which could be problematic.  For example, run-pass options have become increasingly popular in recent seasons, with teams like the 2018 Baltimore Ravens using this as a core feature of their offensive game-plan in the second half of the season \citep{RPO}. One could additionally model this decision at the start of the play, but currently our models condition on the play type at the top level of the framework in Figure \ref{fig:framework}.  

%After we estimate the end-of-play yard line, we can easily determine the additional covariates in the play value model from Section \ref{sec:play-value}.  For example, the updated down number and yards to go depend only on the previous yards to go and the yards gained on the play.  Similarly, the possession team is easily determined, since the pass catcher is either on the offensive or defensive team, and turnovers on downs occur only if the yards gained on the play is less than the previous yards to go. 

%While we focus on continous-time expected end-of-play yard line predictions in this manuscript, we demonstrate in Section \ref{sec:discussion} the use of conditional density estimation to model the distribution of predicted end-of-play yard line values. 
%\todo{LR: The only obvious thing missing from the framework is fumbles and interceptions. Maybe we should write a line on them?  SV:  Will put in discussion section.}
%\todo{RY: We should just say we're implemented the ball-carrier and that others are future work, it's repetitive to say it for each model}

%$EP_{i,t} = $
%Combining the yard line distributions $g(Y_i|\mathscr{F}_{carry}(X_{t,i}))$ across each possible $(d_k, r_j)$ gives us a final distribution on $Y_i$, the yard line at the end of a play (and, consequently, updated values for the down, yards to go, etc)

%The is plugged into the chosen play value model from Section \ref{sec:play-value}, along with reasonable choices for other covariates (time remaining, possession team, score differential, etc) at the end of the play.
    
\subsection{Ball-Carrier Model}
\label{sec:ball-carrier-intro}

First, we model $\mathbb{E}[Y_i|\mathscr{F}(X_{t,i})]$, the expected yards gained by the ball-carrier from their current position on the field conditional on the team in possession and the locations and trajectories of all twenty-two players on the field (including the ball-carrier).

This ball-carrier model is the most important model in our continuous-time play value framework, because (1) it is the only model used for all rushing plays, and (2) all dropback plays that result in a completed pass, an interception, or a scramble/sack require the estimation of the yards gained by the ball-carrier (e.g. the receiver after catching the ball, defender after intercepting the ball, or the QB) from the current position on the field.

Of key importance, only a single model is needed, and this model can be used for \emph{any} situation in which a player is carrying the ball (with no intent to pass).  In other words, our framework requires a single model for all of the following ball-carrier situations:

\begin{itemize}
    \item skill position players (running backs, quarterbacks, wide receivers, etc) on rushing plays,
    \item quarterback on scrambles or sacks,
    \item pass-catcher after that player catches the ball (comprising both offensive players who catch a pass and defensive players who intercept a pass).
\end{itemize}

We experiment with several implementations of this model for rushing plays, described in Section \ref{sec:ball-carrier}. Once we estimate $\mathbb{E}[Y_i|\mathscr{F}(X_{t,i})]$, we can easily obtain the expected end-of-play yard line, $\mathbb{E}[Y_i^*|\mathscr{F}(X_{t,i})]$, by adding the ball-carrier's current yard line to $\mathbb{E}[Y_i|\mathscr{F}(X_{t,i})]$, due to linearity of expectations.

\subsection{Quarterback Decision Model}
\label{sec:qb-decision}

For dropbacks, we must model the decision that a quarterback will make.  Specifically, on a given passing play, the quarterback has three possible decisions:

\begin{itemize}
    \item $d_{ta}$:  throw the ball away,
    \item $d_{r}$:  scramble or be sacked,
    \item $d_{p}$:  pass to a receiver.
\end{itemize}

Let $P(D_i|\mathscr{F}(X_{t,i}))$ be a probability mass function for the decision made by the quarterback on play $i$, a dropback, conditional on the locations and trajectories of all players and the ball over the course of play $i$ up until time $t$.  $P(D_i|\mathscr{F}(X_{t,i}))$ follows a multinomial distribution over the set $\{d_{ta}, d_{r}, d_{p} \}$.   

Possible methods for implementing this model include recurrent neural networks with a multinomial response, multinomial logistic regression, or decision tree frameworks like random forests \citep{Breiman01} or XGBoost \citep{Chen16}. 

\subsection{Pass Target Probability Model}
\label{sec:target-prob}

For passing plays where the QB's decision is to pass (rather than run, be sacked, or throw the ball away), we must model each receiver's target probability, $P(T_{j,i} = 1|\mathscr{F}(X_{t,i}), D_i = \mbox{Pass})$.  Since $T_{j,i}$ is a binary response variable, there are many suitable methods implementing this model.  

Importantly, when training this model, each play in the tracking dataset should be replicated five times (once for each possible targeted receiver on the offensive team), and each replicated play's explanatory and response variables should be updated to be with respect to the receiver in question.  That is, if a receiver $j_1$ is targeted on this play, then $T_{j_1} = 1$, and $T_{j_2} = T_{j_3} = T_{j_4} = T_{j_5} = 0$.  Similarly, $\mathscr{F}(X_{t,i})$ will be with respect to $j_1$.

Once the target probability is calculated for each of the five receivers, these five quantities must be Softmax-normalized so that they form a valid probability distribution over the space of possible targeted receivers.

\subsection{Global Catch Probability Model}
\label{sec:incompletion-model}

For each possible targeted receiver, we next model $P(C_i = 1 | \mathscr{F}(X_{t,i}), D_i = \mbox{Pass}, T_j = 1)$, the probability that a pass to that receiver will be caught by anyone, either the five eligible offensive receivers or eleven defensive players. We first model the \textit{global} catch probability so that the catch probabilities for each offensive receiver and defensive player (from the subsequent \textit{individual} catch probability model) can be computed with the same model, and then Softmax-normalized to the quantity $P(C_i = 1 | \mathscr{F}(X_{t,i}), D_i = \mbox{Pass}, T_j = 1)$.

A pass can only be caught or not caught, so our random variable $C_i$ can take only two values:  1 if the pass is caught (reception or interception), and 0 if the pass is not caught (incomplete).  Since $C_{i}$ is a binary response variable, there are many suitable methods implementing the \textit{global} catch probability model (e.g. logistic regression, tree-based methods, or a recurrent neural network with a binomial response).

\subsection{Individual Catch Probability Model}
\label{sec:catch-model}

Finally, we model $P(C_{k,i} = 1 | \mathscr{F}(X_{t,i}), D_i = \mbox{Pass}, T_j = 1, C_i = 1)$, the probability that player $k$ catches the ball, given that the pass targeted to receiver $j$ was caught.

Similar to the target probability model, when training the catch probability model, each play in the tracking dataset should be replicated sixteen times (once for each of the five eligible receivers on the offensive team, and once for each of the 11 defensive players), and each replicated play's explanatory and response variables should be updated to be with respect to the receiver in question.  That is, if a receiver $k_1$ is targeted on this play, then $C_{k_1,i} = 1$, and $C_{k_2,i} = ... = C_{k_{16},i} = 0$.  Similarly, $\mathscr{F}(X_{t,i})$ will be with respect to $k_1$.

Once the catch probability is calculated for each of the sixteen possible pass-catchers, these sixteen quantities must be Softmax-normalized so that they form a valid probability distribution over the space of possible pass-catchers.

Since $C_{k,i}$ is a binary response variable, there are many suitable methods implementing the incompletion model (e.g. logistic regression, tree-based methods, or a recurrent neural network with a binomial response). See \citet{Deshpande19} for an extensive study of this problem.

%After we estimate the end-of-play yard line, we can easily determine the additional covariates in the play value model from Section \ref{sec:play-value}.  For example, the updated down number and yards to go depend only on the previous yards to go and the yards gained on the play.  Similarly, the possession team is easily determined, since the pass catcher is either on the offensive or defensive team, and turnovers on downs occur only if the yards gained on the play is less than the previous yards to go. 

%While we focus on continous-time expected end-of-play yard line predictions in this manuscript, we demonstrate in Section \ref{sec:discussion} the use of conditional density estimation to model the distribution of predicted end-of-play yard line values. 

\section{The Ball-Carrier Model}
\label{sec:ball-carrier}

An advantage of our framework is the modularity of the models. Modularity implies that we can develop each model independently, then plug the best model for each task into the framework. For example, once we develop a ball-carrier model, we can use this model to compute continuous-time expected end-of-play yard line predictions for each moment in a game when a player is running with the football.

Our ball-carrier model estimates the expected yards gained from the player's current yard line (and thus the final yard line a ball carrier will reach on a play), conditional on the locations and trajectories of all twenty-two players on the field. 
Formally, a play with $L$ frames will generate $L$ ``observations'' for the model's training. 
For example, the player's trajectories up to frame $1\le l \le L$ will form a single observation, where the dependent variable will be the yards the rusher gained from that point on. 

Section \ref{sec:filtrations} introduces the features we use for our ball-carrier model, Section \ref{sec:models} describes the different ball-carrier models we tried, and Section \ref{sec:model-val} describes how we evaluate our ball-carrier models.

\subsection{Features for the Ball-Carrier Model}
\label{sec:filtrations}
The tracking data provides a wealth of information about a football play, including who is on the field, where they are on the field, which direction they are facing, how fast they're running, and more.  A first step in developing our ball-carrier model is deciding what information will be helpful to use in modeling the yards gained from the ball-carrier's current position. Since the ball-carrier model within our framework is meant to work in general ball-carrier situations (running plays, QB scrambles, receivers after catching a pass, etc), we consider a broad class of features that are applicable across all situations. Future work focusing on particular situations, e.g. only running back carries at hand-off, may consider features and player roles inherent to that situation (e.g. personnel packages, men in box).

For the first set of features, rather than simply using the raw $(x,y)$ coordinates and direction of the players, we create \textit{adjusted} versions that are calculated with respect to the ball-carrier's endzone. For each player, we record their: (1) \textit{adjusted} $x$-coordinate denoting how many yards away they are from the ball-carrier's target endzone; (2) \textit{adjusted} $y$-coordinate denoting how far they are from the middle of the field, where positive values denote the left side (with respect to facing the target endzone), and negative values denote the right; and (3) the player's direction with respect to the target endzone, where $0^{\circ}$ is straight towards to the endzone, positive angles denote directions to the left, and negative angles denote directions to the right. 

\begin{figure}
    \centering
    \includegraphics[width= \textwidth]{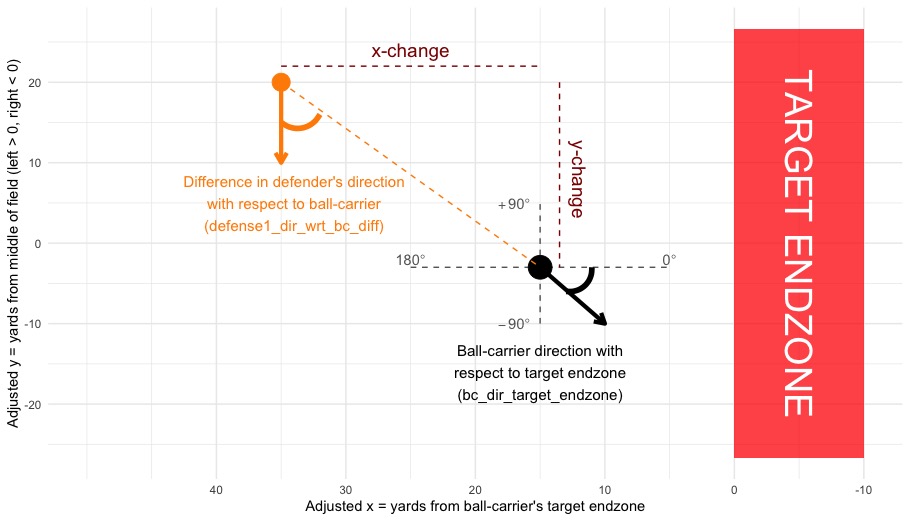}
    \caption{A visual explanation of \textit{adjusted} $x$,$y$ coordinates and direction features, where the orange point denotes an example defender and the black point is an example ball-carrier. The arrows denote the players' actual directions. The gray reference lines display the ball-carrier's direction with respect to the target endzone. The orange reference line displays the defender's direction with respect to the ball-carrier.  The dashed darkred lines denote the $x$,$y$-change features.}
    \label{fig:adj_xydir_features}
\end{figure}

The next set of features is based on the location and direction of each player relative to the ball-carrier. For each player, we record their Euclidean distance to the ball-carrier, as well as the difference between the ball-carrier's \textit{adjusted} $x$-coordinate ($y$-coordinate) with the player's \textit{adjusted} $x$-coordinate ($y$-coordinate), preserving the $x$ (or $y$) \textit{change} between the player and ball-carrier in one-dimension while maintaining the orientation of the field with respect to the target endzone. The resulting $x$-change variable denotes how far away the player is to the ball-carrier, where $x$-change $>0$ indicates the player is closer to the endzone while $x$-change $<0$ means the ball-carrier is closer to the endzone. The $y$-change variable captures the vertical location of the player with respect to the ball-carrier, computing the difference between the player's \textit{adjusted} $y$-coordinate and the ball-carrier's; the sign of the $y$-change variable denotes whether the player is above ($y$-change $> 0$) or below ($y$-change $< 0$) of the ball-carrier, or similarly left or right of the ball-carrier respectively with respect to facing the endzone. For defensive players, we additionally compute the absolute difference between the defender's direction and the angle of the shortest segment between the defender and the ball-carrier. A visual explanation of the features capturing information with respect to the target endzone and the ball-carrier are displayed in Figure \ref{fig:adj_xydir_features}. 

We split the players into three groups: ball-carrier (\texttt{bc\_}), offense (\texttt{offenseX\_}), and defense (\texttt{defenseX\_}), where for the offensive and defensive groups of players, we order the players based their Euclidean distance to the ball-carrier. For example, the feature {\tt defense1\_x\_adj} denotes how many yards away the closest defender is from the ball-carrier's target endzone, while the feature {\tt bc\_s} gives the speed of the ball-carrier, and so on. The full list of features constructed for the players is displayed in Table \ref{tab:player-vars}, where \texttt{x\_change}, \texttt{y\_change}, and \texttt{dist\_to\_ball} are for the \texttt{X} closest teammates or defenders only, and \texttt{dir\_wrt\_bc\_diff} is for the \texttt{X} closest defenders only.

\begin{table}[!ht]
\centering
\caption{List of features constructed for players.}
\label{tab:player-vars}
\centering
    \begin{tabular}{p{4cm} p{9cm}}
        \hline
        Variables & Description  \\
        \hline
        \texttt{x\_adj}  & Horizontal yards from ball-carrier's target endzone. \\
        \hline
        \texttt{y\_adj}  & Vertical yards from center of field with respect to target endzone, where positive values indicate left side while negative values indicate right side.     \\
        \hline
        \texttt{dir\_target\_endzone} & Direction player is facing with respect to the target endzone, where 0 indicates facing the endzone while positive degrees between 0 and 180 denote facing to the left while negative degrees between 0 and -180 denote facing to the right. \\
        \hline
        \texttt{s}  & Speed in yards/second.   \\
        \hline
        \texttt{dis} & Distance traveled since previous frame.  \\
        \hline
        \texttt{x\_change} & Difference between \texttt{bc\_x\_adj} and \texttt{X} closest offensive or defensive player's \texttt{x\_adj}. \\
        \hline
        \texttt{y\_change} & Difference between \texttt{X} closest offensive or defensive player's \texttt{y\_adj} and \texttt{bc\_y\_adj}. \\
        \hline
        \texttt{dist\_to\_ball} & Euclidean distance between \texttt{X} closest offensive or defensive player and ball-carrier. \\
        \hline
        \texttt{dir\_wrt\_bc\_diff} & Minimal absolute difference between \texttt{X} closest defender's direction and the angle between the defender with the ball-carrier. \\
        \hline
    \end{tabular}
\end{table}

We additionally use each player's speed and distance traveled from the previous frame, along with features extracted from the Voronoi tessellation of player locations \citep{Voronoi1908}. The Voronoi tessellation partitions the playing surface into regions, where each region corresponds to the area of the playing surface closest to an individual player. These regions help expose some of the more complex geometric relationships between the players.\footnote{Several authors use Voronoi tessellations to analyze tracking data in sports, such as \citet{hochstedler2016finding}.  For an overview, see \cite{GudmundssonH16}.} We construct two different versions of Voronoi tessellations: (1) using all twenty-two players and (2) using only the ball-carrier and the eleven defenders. By removing the ball-carrier's teammates from the tessellation, we are attempting to capture blocking information representing the goal of teammates to assist the ball-carrier directly. We created the Voronoi tessellations with the \texttt{deldir} package in \texttt{R} \citep{deldir}.

\begin{figure}
    \centering
    \includegraphics[width= \textwidth]{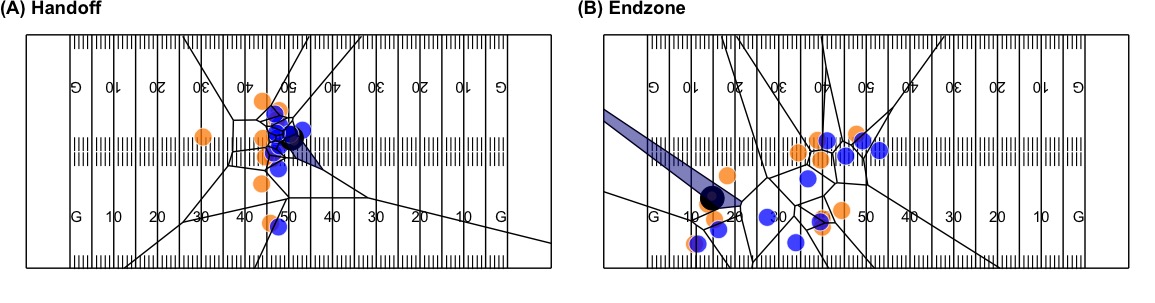}
    \caption{A display of the Voronoi tessellation for Cordarrelle Patterson's 47-yard TD run at (A) handoff and (B) a frame when his region includes the target endzone. The region for the ball-carrier is shaded.}
    \label{fig:exvoronoi}
\end{figure}

We extract simple features from these Voronoi tessellations: (1) the total area of the ball-carrier's Voronoi tessellation, (2) the area of the ball-carrier's tessellation between them and the target endzone, and the $x$-coordinates of the (3) closest and (4) farthest points from the target endzone on the boundary of the ball-carrier's Voronoi region. For the tessellation constructed using the ball-carrier's teammates, we additionally create an indicator variable determining whether or not all edges of the ball-carrier's tessellation are shared with their teammates. Table \ref{tab:voronoi-vars} lists the variables constructed from the Voronoi tessellations including all players, where we denote the four additional variables from Voronoi tessellation with the ball-carrier's teammates removed using \texttt{bc\_only} (e.g. \texttt{voronoi\_bc\_only\_area}). In total there are nine variables extracted from Voronoi tessellations. Figure \ref{fig:exvoronoi} displays examples of the tessellations with all players accounted for displaying a shaded area for the ball-carrier's region at (A) handoff and (B) a point in the run when the ball-carrier's tessellation includes the target endzone. 

\begin{table}[!ht]
\centering
\caption{List of features constructed from Voronoi tessellations with all players.}
\label{tab:voronoi-vars}
\centering
    \begin{tabular}{p{5cm} p{8cm}}
        \hline
        Variables & Description  \\
        \hline
        \texttt{voronoi\_bc\_close\_adj} & Horizontal yards away from ball-carrier's target endzone for the point on the perimeter of the ball-carrier's Voronoi region that is closest to the target endzone. \\
        \hline 
        \texttt{voronoi\_bc\_far\_adj} & Horizontal yards away from ball-carrier's target endzone for the point on the perimeter of the ball-carrier's Voronoi region that is farthest to the target endzone. \\
        \hline 
        \texttt{voronoi\_bc\_area} & Total area of the ball-carrier's Voronoi region. \\
        \hline 
        \texttt{voronoi\_bc\_area\_in\_front} & Total area of the ball-carrier's Voronoi region between the ball-carrier and the target endzone. \\
        \hline 
        \texttt{voronoi\_bc\_bubble} & Indicator if all edges of the ball-carrier's Voronoi region are shared with their teammates. \\
        \hline 
    \end{tabular}
\end{table}

We consider of all the features (centered and scaled) in Tables \ref{tab:player-vars} and \ref{tab:voronoi-vars} for the ball-carrier model. We also explored lagged versions of the variables, but did not find that these variables improved the performance of the models considered in Section \ref{sec:models}. We emphasize that this set of features is only a starting point and future feature engineering, such as the space ownership approach from \cite{Fernandez18}, may significantly improve the ball-carrier model's performance. Similarly, we currently are lacking an approach for directly accounting for the positioning of blockers, which may be especially useful for ball-carrier segments in the open field (though this is done indirectly via the Voronoi features). Improving upon the feature space used as input for the ball-carrier model may improve the model's prediction accuracy, and is a task left to future work.

\subsection{Models}
\label{sec:models}
The ball-carrier model has several important aspects:

\begin{itemize}
    \item {\bf High dimensions}. Since there are twenty-two players on the field, and each player has an $x$-coordinate, $y$-coordinate, direction, speed, etc., we can use many features to estimate the final yard-line of the ball-carrier. 
    \item {\bf Non-linearity}. We don't expect the best prediction for the final yard-line to have a simple linear structure. For example, we would expect a player facing the ball-carrier to have a better chance of making the tackle than a player than a player not facing the ball-carrier.
    \item {\bf Interactions}. Our features should depend on each-other, for example the speed and direction the ball-carrier is facing with respect to the target endzone.
    \item {\bf Time}. Since we're estimating the final yard-line at each time frame, the predictions should be smooth from frame-to-frame, and we should be able to use this temporal structure in our models.
\end{itemize}

Thus, we select models that capture these aspects of the data, and we use appropriate regularization to avoid overfitting. Before moving to more complicated models, we establish a baseline \textit{intercept-only} model that doesn't use any of the features described in Section \ref{sec:filtrations}. We use the intercept-only model to set an initial performance benchmark.

The next model we consider is the LASSO regression model \citep{Tibshirani94}. The LASSO works well in high dimensions, is easy to interpret, and has a fast implementation. We use the \texttt{glmnet} implementation in \texttt{R}, choosing the one standard error regularization penalty from model training via cross validation \citep{glmnet}. Since this is a linear model, we make an additional adjustment to the direction and $y$-coordinate based variables, using the absolute value (ie how far from middle of field) rather than allowing the sign to dictate which side of the field a player is on or facing. This is an obvious limitation of the LASSO model that would require further additional feature engineering to compare in performance to nonlinear models.

We also explored gradient boosted trees using the popular XGBoost implementation \citep{Chen16}. Like the LASSO, XGBoost works in high dimensions, but also accounts for non-linear interactions in the data via tree-based partitioning. Of course, the LASSO can also account for non-linear interactions, but that would require the explicit construction of additional features. We implemented XGBoost via the \texttt{xgboost} \texttt{R} package, and found the default settings (100 trees, max depth of 3 splits) to yield the best results in cross-validation (CV) training among the complexity parameters that were considered.

Another flexible model that works well in high dimensions, and can capture non-linear interactions, is a feedforward neural network \citep{Haykin98}.  Chapter 6 of \cite{Goodfellow-et-al-2016} provides a clear and detailed overview of this type of model. Based on CV results across a number of layers and hidden units, we used a feedforward neural network with two layers, where each layer has fifty hidden units. We use a ReLu activation function for each layer\footnote{\cite{ReLu} describe the ReLu activation function, and show that it outperforms other activation functions for deep networks.}, and regularized each layer with an L1 penalty.  We trained the network with the Adam algorithm \citep{kingma2014adam}, and implemented the network using the {\tt keras} {\tt R} package \citep{keras}.

So far, none of our models have explicitly accounted for the temporal structure of the data. To remedy this, we can adapt our feedforward neural network into a {\it recurrent neural network}. Specifically, we use a long short-term memory (LSTM) network \citep{hochreiter1997long}. Our LSTM has two layers with fifty inputs each layer (determined by CV tuning), and we use a recurrent dropout rate of 20\% for each layer. Finally, because not all ball-carrier sequences are the same length, we zero-pad each sequence to the size of the longest ball-carrier sequence. Table \ref{tab:models} summarizes the five different models we use, in terms of aspects we considered in the beginning of this section.

\begin{table}[!ht]
\centering
\caption{Comparison of ball-carrier models.}
\vspace{1em}
\label{tab:models}
\centering
\scalebox{0.8}{
\begin{tabular}{|p{6cm} | p{3cm} | p{2cm} | p{2cm} | p{1cm} |}
    \hline
    {\bf Model} & High-dimensions & Non-linear & Interactions & Time  \\
    \hline
    Intercept-only (baseline) & & & & \\    
    \hline
    LASSO & \checkmark & & & \\    
    \hline
    XGBoost & \checkmark & \checkmark & \checkmark & \\    
    \hline
    Feedforward Neural Network & \checkmark & \checkmark & \checkmark & \\   
    \hline
    Long short-term memory (LSTM) & \checkmark & \checkmark & \checkmark & \checkmark \\    
    \hline
\end{tabular}}
\end{table}

\subsection{Model Validation}
\label{sec:model-val}

Since our goal is to generate continuous-time end-of-play yard line predictions for every player-tracking frame in the data, we need to ensure that our selected model is performing well across the sample of provided games. As a computationally feasible alternative to leave-one-play-out CV, we use leave-one-week-out (LOWO) CV (e.g. train on all frames from games in weeks one through five, then generate predictions on all frames from games in holdout week six) to select the ball-carrier model. We evaluate the LOWO predictions with two criteria: (1) overall root mean-squared error (RMSE), and (2) both the RMSE and residuals across the number of frames from start of ball-carrier sequence.

The first criterion, overall holdout RMSE, is connected to our goal of generating baseline continuous-time within-play values across all individual frames. The second criteria is to ensure our model is performing well across the entire ball-carrier sequence, with no systematic bias at certain points during the course of the run. A model is unlikely to accurately forecast the outcome of a ball-carrier sequence at the first frame (i.e. handoff) when the ball-carrier sequence is long. However, the model should be more accurate as the run progresses given the observed filtration of information.

\section{Results}
\label{sec:results}
This section walks through various results and analysis of our ball-carrier model. 

\subsection{Model Comparison and Selection}
\label{sec:selection}

\begin{figure}[!ht]
    \centering
    \includegraphics[width= \textwidth]{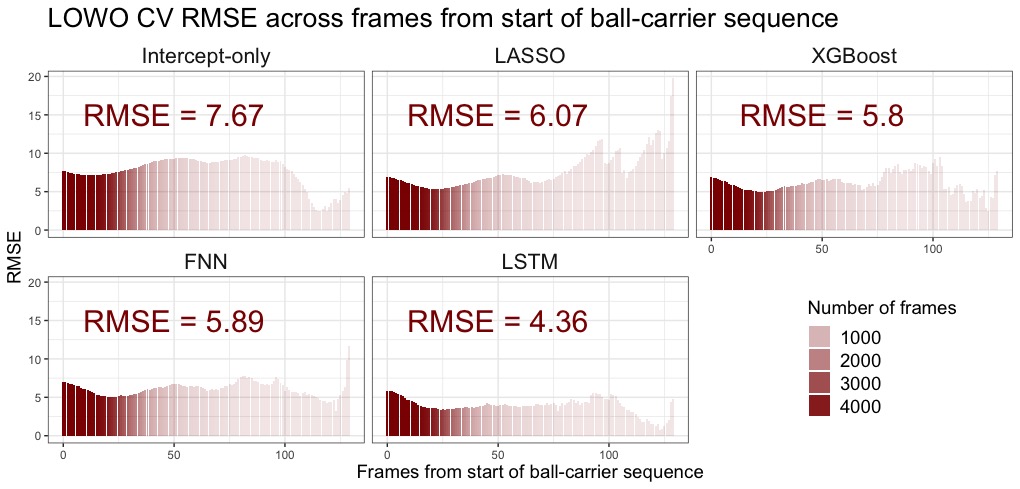}
    \caption{Comparison of LOWO CV RMSE values for each model by number of frames from start of ball-carrier sequence, with overall LOWO CV RMSE values displayed in text.}
    \label{fig:lowo-cv-rmse}
\end{figure}

Figure \ref{fig:lowo-cv-rmse} displays both the overall LOWO CV RMSE for each model in text, and visually the RMSE across the ball-carrier sequence by the number of frames from start (i.e. zero frames away corresponds to the handoff / start of run). Unsurprisingly, all covariate-informed models perform better than the intercept-only baseline. We can clearly see that the LSTM model outperforms other approaches with regards to overall error as well as the lowest RMSE across the entire ball-carrier sequence.

\begin{figure}[!ht]
    \centering
    \includegraphics[width= \textwidth]{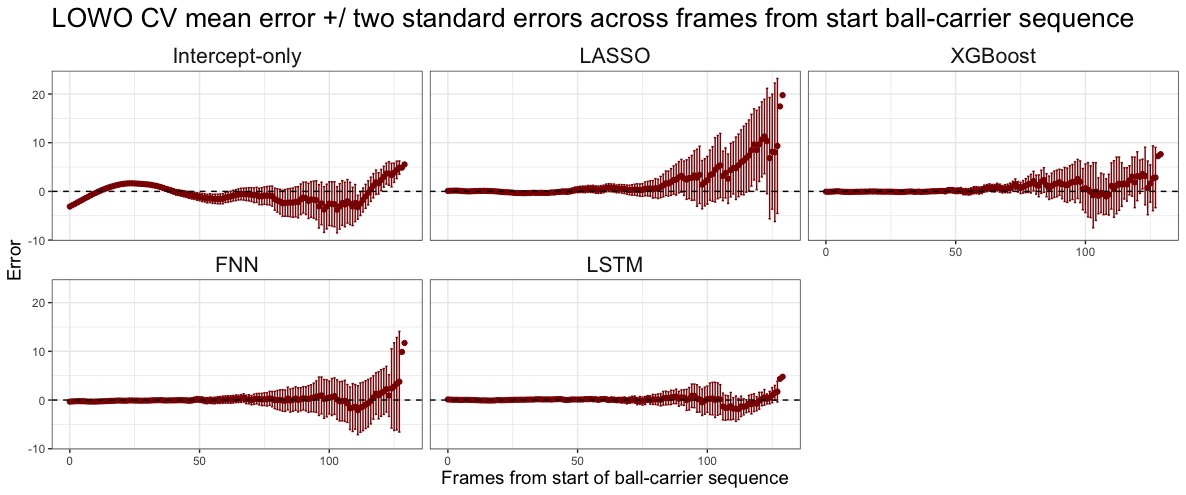}
    \caption{Comparison of LOWO CV mean error values (denoted by points) with plus/minus two-standard errors for each model by number of frames from start of ball-carrier sequence.}
    \label{fig:lowo-cv-error}
\end{figure}

Figure \ref{fig:lowo-cv-error} displays the LOWO CV mean error with plus/minus two standard errors for each model by number of frames from start of ball-carrier sequence. We can clearly see the temporal bias of the intercept-only baseline, and that the LSTM model clearly makes the smallest long-term errors. Since the LSTM is the best-performing model based on the LOWO CV criteria considered, we use its predicted values for the example play and player evaluations presented in Section \ref{sec:ep-players}.

\subsection{Analysis of Feature Importance}

\begin{figure}[!ht]
    \centering
    \includegraphics[width= \textwidth]{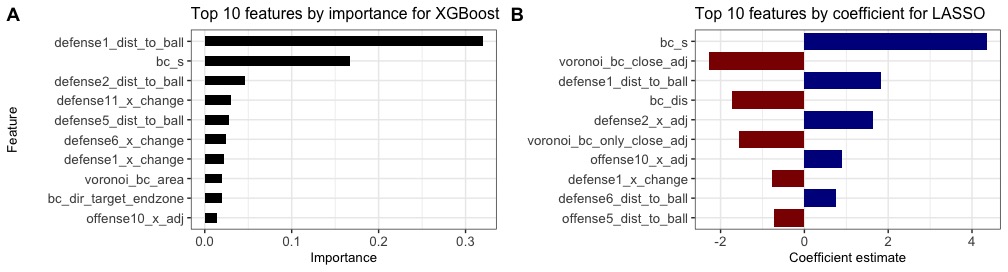}
    \caption{(A) Top ten features by importance for the XGBoost model trained on all data. (B) Top ten features by absolute value of coefficient estimates for LASSO trained on all data (blue denotes positive, red denotes negative coefficient values).}
    \label{fig:var-plots}
\end{figure}

For context regarding the covariates considered, we additionally trained the XGBoost and LASSO models on the entire dataset. Figure \ref{fig:var-plots}(A) displays the top ten features by importance from the XGBoost model. The two most important features are the distance the to closest defender (\texttt{defense1\_dist\_to\_ball}) and the ball-carrier's current speed (\texttt{bc\_s}). This is consistent with the top variables selected by the LASSO model trained on the entire dataset, as displayed in Figure \ref{fig:var-plots}(B). The directions of the LASSO coefficients are consistent with intuition, e.g. the faster the ball-carrier is moving the further they are expected to carry the football. We also see that information extracted from Voronoi tesselations such as the ball-carrier area, as well as the points closest to the target endzone, are selected by the LASSO model. Based on the LASSO coefficients, the farther away from the endzone the closest Voronoi point to the target endzone gets - we expect to see a decrease in the number of yards gained, which matches intuition.

\subsection{Continuous-Time Prediction Examples}
\label{sec:ep-play-ex}

\begin{figure}
    \centering
    \includegraphics[width= \textwidth]{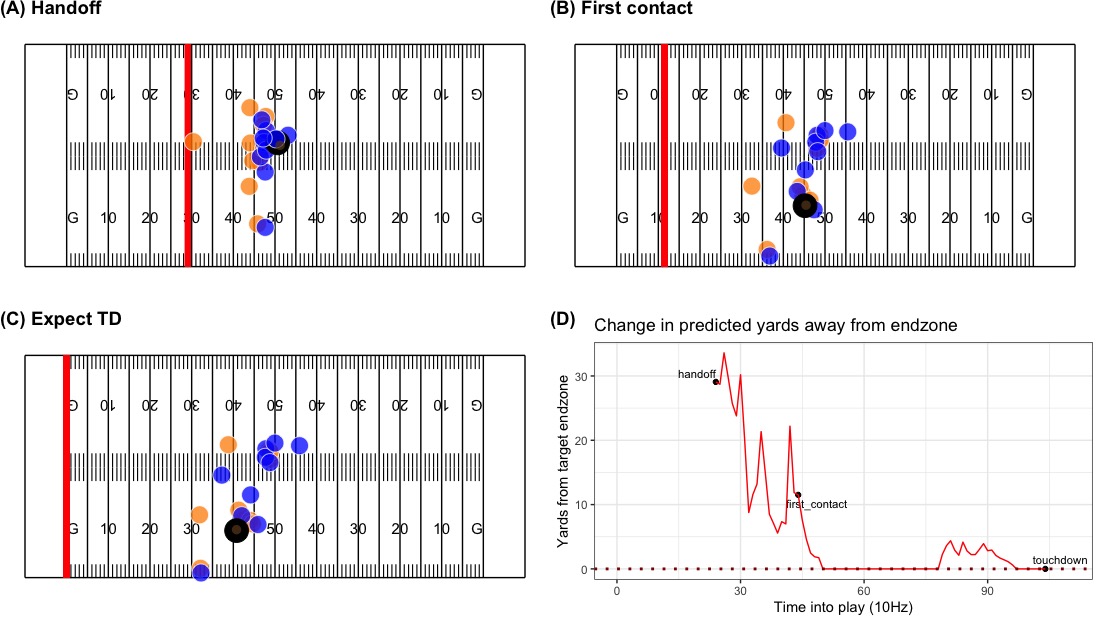}
    \caption{The red line indicates the expected yard line Cordarrelle Patterson (in black) will reach at (A) handoff, (B) first contact, and (C) the first frame he's expected to reach the endzone. Blue points indicate the ball-carrier's teammates while orange represents the opponents. (D) Predicted yards from target endzone over the course of the 47-yard TD run.}
    \label{fig:run-pred-ex}
\end{figure}

\autoref{fig:run-pred-ex} displays an updated version of \autoref{fig:explay} with the expected yard line (in red), that the ball-carrier (black) is predicted to reach given all information regarding his teammates (blue) and opponents (orange) using the LSTM model at (A) handoff, (B) first contact, and (C) the first frame when the expectation was a touchdown. Figure \ref{fig:run-pred-ex}(D) displays the change in the predicted yards from the target endzone over the course of the 47-yard TD run. At handoff, the expectation is roughly a 15-yard gain and displays sharp changes in predictions with the noticeable gain following the point of first contact where the ball-carrier is then expected to reach the endzone.

\begin{figure}[!ht]
    \centering
    \includegraphics[width= \textwidth]{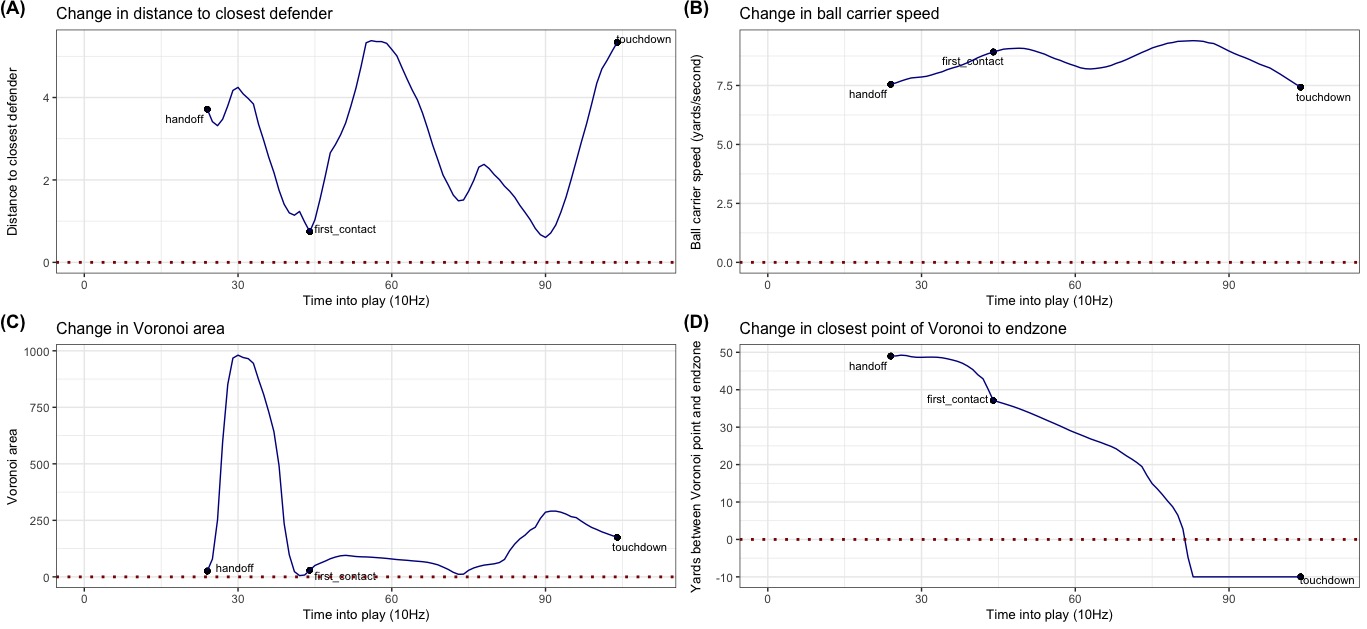}
    \caption{The change in (A) distance to closest defender, (B) ball-carrier speed, (C) the ball-carrier's Voronoi area during, and (D) the yards between the closest point of Patterson's Voronoi and the target endzone.}
    \label{fig:run-pred-ex-vars}
\end{figure}

For context in understanding the change in predicted yards gained throughout Patterson's touchdown run, \autoref{fig:run-pred-ex-vars} displays the (A) change in the distance to closest defender, as well as (B) Patterson's speed, (C) his Voronoi area and (D) the yards between the closest point of his Voronoi and the target endzone in each frame of the run. We see that the moment Patterson was no longer expected to score a touchdown occurred when the closest defender was within the same distance as the point of first contact. But he then gained additional separation from the opponent, leading to an expectation of scoring a touchdown once again. We clearly see that using the Voronoi alone is insufficient in predicting the yards gained, as Patterson maintains a high level of speed throughout his run. While our LSTM model accounts for the complex relationships between the considered features in Section \ref{sec:filtrations}, future work should explore features extracted from weighted Voronoi tessellations \citep{cervone2016nba}.

\subsection{Player Evaluation Examples}
\label{sec:ep-players}

\begin{figure}
    \centering
    \includegraphics[width= \textwidth]{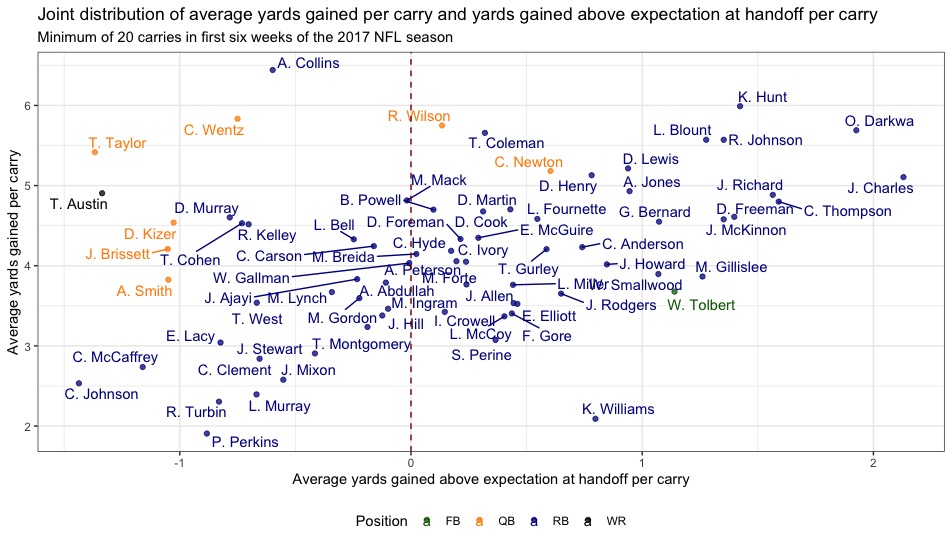}
    \caption{Joint distribution of average yards gained per carry and yards gained above expectation at handoff per carry for players with at least twenty carries.}
    \label{fig:player-handoff-chart}
\end{figure}

As noted in Section \ref{sec:prev-football}, we can use the resulting continuous-time predicted yardline from the LSTM model to gain insight into the contributions of individual athletes over the course of a play. For instance, Figure \ref{fig:player-handoff-chart} displays the joint distribution of average yards gained per carry and yards gained above expectation at handoff per carry for players with at least twenty carries in the available sample of tracking data. We see clear divergences between players and their observed yards gained per carry and how many the player has gained above expectation at handoff. For instance, Alex Collins is leading ball-carriers in average yards gained per carry in the sample, but on average has gained negative yards relative to expectation to handoff. 

Since our model provides continuous-time predictions, we examine the joint distribution of the same set of players' average yards gained above expectation per carry at handoff and one second into the carry. We examine one second (or ten frames) into the carry to account for the change in the positioning of the offense's blocking scheme. The horizontal and vertical dashed lines separate the players that are over/under performing at the respective points in time. We see Le'Veon Bell under-performed with respect to expectation at handoff but over-performed expectation at one second into the carry.

\begin{figure}
    \centering
    \includegraphics[width= \textwidth]{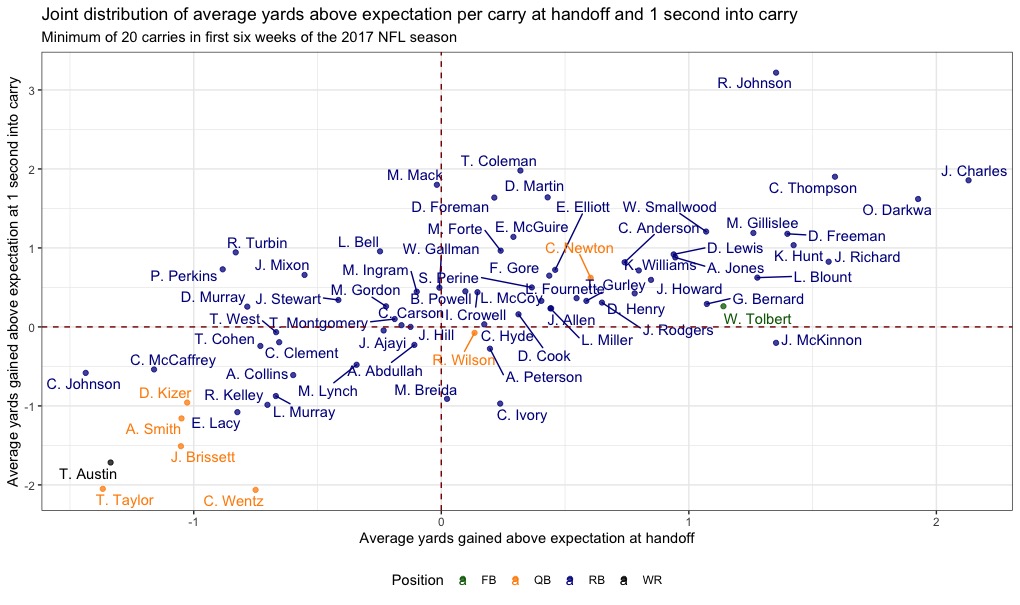}
    \caption{Joint distribution of average yards gained above expectation per carry at handoff and one second into carry for players with at least twenty carries.}
    \label{fig:player-second-chart}
\end{figure}

With limited data, it is difficult to evaluate these continous-time metrics and make claims about their stability and discriminatory ability. Each of our ball-carrier estimates are a function of all twenty-two players on the field, while the above metrics are merely attributing the observed change in value of the frame-level data to the ball carrier. Regression-based approaches such as the implementation in \citet{Yurko19} could provide a starting point for dividing the credit among players within the play. Additionally, our model accounts for the player's speed as an input which is an inherent function of the ball-carrier. Future work would consider imputing average speed levels for all ball-carriers at particular moments over the course of the run or generate the ball-carrier model without speed accounted for. However, due to the limited availability of data this currently presents a challenge that could be addressed when more data are made available.

\section{Discussion \& Future Directions}
\label{sec:discussion}

In this work, we provide a framework for continuous-time within-play valuations of game outcomes in football using player and ball-tracking data from the NFL.  We implement the core piece of this framework, a model for the expected yards gained from a ball-carrier's current yard line, conditional on the locations and trajectories of all twenty-two players on the field, and we test several different modeling approaches for doing so.  As input for this ball-carrier model, we create a rich set of features that describe the location of the ball-carrier relative to the target endzone and other players on the field, e.g. with features generated from Voronoi tessellations.  For this ball-carrier model, we find that all tested models substantially outperform a baseline intercept-only model, but that a long short-term memory (LSTM) recurrent neural network outperforms alternative approaches according to the LOWO CV evaluation measures we set forth in this paper. Finally, we provide examples continuous-time predictions and player evaluation metrics using the 
NFL-provided tracking data from the first six weeks of the 2017 regular season.

\subsection{Conditional Density Estimation}

\begin{figure}[!ht]
    \centering
    \includegraphics[width= \textwidth]{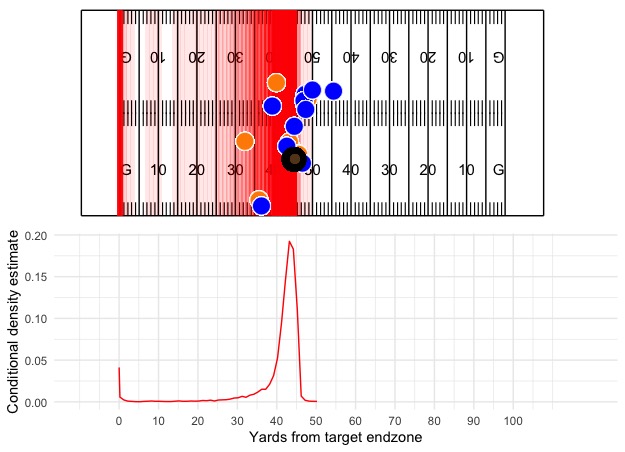}    
    \caption{Conditional density estimate of end-of-play yard line at point of first contact. The opacity of the red lines denotes the estimated density for each yard lines (darker = higher density, lighter = lower density), with the corresponding density values displayed in the curve at the bottom.}
    \label{fig:rfcde-contact-cuve}
\end{figure}

Although we introduced our framework in Section \ref{sec:framework} with respect to predicting the expected end-of-play yard line, $\mathbb{E}[Y_{t,i}^* | \mathscr{F}(X_{t,i})]$, we can replace our ball-carrier model in Section \ref{sec:ball-carrier-intro} with an estimate for the conditional density function of the end-of-play yard line $\hat{f}(Y_{t,i}^* | \mathscr{F}(X_{t,i}))$. Rather than make parametric assumptions, we demonstrate the use of random forests for conditional density estimation (RFCDE) as a flexible approach that allows us to incorporate all of the features covered in Section \ref{sec:filtrations} \citep{RFCDE}. This results in a estimate for an entire density curve over each potential yard line, as displayed for example in Figure \ref{fig:rfcde-contact-cuve} at the point of first contact during the example touchdown run by Patterson revealing a bi-modal distribution, with the model predicting Patterson's more likely to end near his current location but a potential for reaching the endzone.

Let $g(Y_{t,i}^* | \mathscr{F}_{carry}(X_{t,i}))$ represent an arbitrary function for play value, such as the EP or WP models from \cite{Yurko19} described in Section \ref{sec:prev-play}. We denote this as a function of the end-of-play yard line $Y_{t,i}^*$ for simplicity, since the additional covariates (e.g. down, yards to go, possession team) can be easily determined based on the end-of-play yard line value. However, if we use the estimates for the expected end-of-play yard line,  $\mathbb{E}[Y_{t,i}^* | \mathscr{F}_{carry}(X_{t,i})]$, from our ball-carrier model in Section \ref{sec:ball-carrier} and plug it in as input for $g$, we've simply generated point estimates for the within-play value function. Using these point estimates as input into the non-linear, complex functions for within-play EP and WP can lead to biased or incorrect estimates of the within-play expectation of EP or WP.

Within the context of our framework, we can instead use the full conditional density estimate for the end-of-play yard line $\hat{f}(Y_{t,i}^* | \mathscr{F}(X_{t,i}))$ to compute the expected within-play value,
\begin{equation}
\label{eq:rfcde-ep}
    \mathbb{E}[g(Y_{t,i}^* | \mathscr{F}(X_{t,i}))] = \int g(Y_{t,i}^* | \mathscr{F}(X_{t,i})) \cdot \hat{f}(Y_{t,i}^* | \mathscr{F}(X_{t,i})) \cdot d X_{t,i}.
\end{equation}

The above approach implicitly assumes the independence of the between-play estimates of play value ($g(Y_{t,i}^* | \mathscr{F}(X_{t,i}))$; e.g. EP or WP) and the within-play conditional density estimates of the expected end of play yard line ($\hat{f}(Y_{t,i}^* | \mathscr{F}(X_{t,i}))$).  Since the between-play value models that we use only account for factors that are observable between plays (e.g. down, yards to go, yard line, time remaining, timeouts remaining, etc), these models are intuitively independent of the within-play conditional density estimate of the end-of-play yard line (which only account for information observable within plays).  Researchers looking to replicate our work should be careful about the interplay of these two classes of models.  For example, if one were to account for timeouts remaining in the within-play models for the end of play yard line (e.g. since timeouts remaining may influence player decision-making in some plays), doing so may render this assumption of independence false, since a WP model typically accounts for timeouts remaining as well.  Similarly, if one built new between-play EP or WP models that accounted for player-level characteristics observable with player-tracking data (e.g. fatigue), this assumption may not hold. %Intuitively, this assumption holds:  The within-play estimates use \emph{only} within-play tracking data, and the between-play estimates do not.  Moreover, we posit that no additional information from the \emph{within-play} tracking data will influence the \emph{between-play} valuations of a football game, regardless of which model for between-play valuation is used (Section \ref{sec:prev-play}). That is, the value of a game situation between when the previous play ends and the next play begins is a function of \emph{only} the factors that are observable between plays (e.g. down, yards to go, yard line, score, time remaining, timeouts remaining, etc); these values are conditionally independent of any information that can be gathered from within-play tracking data. Because of this, it is not necessary to develop new models for between-play value using tracking data.
In our case, we use the EP and WP models from \cite{Yurko19} for this purpose, since they only use information observable between plays.  Additionally, these models are reproducible, publicly available in the \texttt{nflscrapR R} package, well-calibrated, and interpretable in terms of game outcomes. 

To demonstrate how conditional density estimation with RFCDE works for this purpose, we perform the integration in Equation \ref{eq:rfcde-ep} with a RFCDE implementation of the ball-carrier model to compute the continuous-time within-play expectations for the \citet{Yurko19} EP and WP estimates.  An example play shown in Figure \ref{fig:ep-wp-example}.  Future work should build upon this proof of concept, and perform the same type of extensive evaluation demonstrated in this manuscript for modeling the expected end-of-play yard line.

\subsection{Future Work}

There are many additional potential directions for future work.  First, we currently do not handle special teams.  A brief sketch of how this important piece of a football may fit into our framework is as follows:  For kickoff and punt returns, we can use the ball-carrier model, provided enough training data (this was not possible with only six weeks of data for this paper).  For field goals, since blocked kicks are rare, continuous-time play value is likely of limited additional value above what is possible with discrete-time (between-play) play value models.  Similarly, blocked punts are rare, so attempting to model these may prove more challenging than its worth.

Second, we currently do not handle fumbles by the ball-carrier.  To do so, we would have to incorporate a survival component into our model, accounting for the hazard of a fumble at each moment throughout a ball-carrier sequence, conditional on the features of that sequence that may be indicative of changes in fumble rates.  However, fumbles are rare events, and even rarer in a six-week sample of games, rendering the estimation of this component of the ball-carrier model impractical.  This task is left to future work, if/when multiple seasons of tracking data are available.

Third, we currently use the observed time of the ball-carrier sequence for adjusting the amount of game time remaining for plugging in as an input to compute the EP and WP estimates displayed in Figure \ref{fig:ep-wp-example}. An elegant approach would be to model the joint distribution of the yards gained from the ball-carrier's current position and the time remaining at the end of the play, which is possible through the use of RFCDE \citep{RFCDE}.  However, doing so would (at least) double the size of the parameter space. Additionally, time remaining is typically of little value in a between-play model for play value, and only comes into play in somewhat rare situations at the end of the 1st or 2nd half.  With a limited set of six weeks of tracking data, the ad hoc approach we use here will suffice.

Fourth, there is more work to be done in the area of feature engineering.  As discussed, using a Voronoi-like approach that accounts for the velocity of players on the field, similar to what \cite{Fernandez18} do for modeling space creation and occupation in soccer, may yield some improvements in model predictions.  Additionally, accounting for blockers (e.g. by joining the adjacent Voronoi polygons of teammates to identify a path through which the ball-carrier can travel) may also lead to improved prediction accuracy.

Fifth, in the context of player evaluation, researchers should be careful about how they use our models when evaluating players.  As demonstrated in Figure \ref{fig:var-plots}, the ball-carrier speed is one of the most important features in modeling yards gained from the current position on the field.  However, if we condition on the speed of a player in the model, any gains a ball-carrier makes as a result of being faster than other ball-carriers (or losses from being slower) will be not be attributed to that ball-carrier.  As such, researchers using our models for player evaluation should consider using the average speed of player when evaluating individuals, so that deviations above and below average are attributed to that player.

Along these lines, future researcher may use our continuous-time, within-play valuation of game outcomes to evaluate micro-actions of all players on the field, similar to what has been done in basketball \citep{Sicilia19} and soccer \citep{Fernandez18, Decroos19}.  Similar ideas have been implemented for players at offensive skill positions at the discrete-time level in football \citep{Yurko19}, but never implemented for all twenty-two players on the field and in a continuous-time framework.

Finally and most importantly, we currently only provide an implementation of the ball-carrier model, and we do not implement the other modular sub-models in our framework for continuous-time play value (e.g. QB decision model, target probability model, catch probability model, etc).  Implementation of these models is somewhat straightforward, given an appropriate feature space:  Since the responses in these models are either binary (target probability, global or individual catch probability) or multinomial (QB decision), simple adjustments can be made to LSTM we use for the ball-carrier model to enable a similar approach to be used for these pieces of the framework.  Additionally, some authors implement excellent versions of these models already.  For example, \cite{Deshpande19} implement a catch probability model, and \cite{Burke19} implements both a QB decision model and a target probability model.  We look forward to incorporating these models in our framework for continuous-time valuation of game outcomes in football.

\bibliographystyle{DeGruyter}
\bibliography{bibliography}

\end{document}